\documentclass[article]{revtex4}

\usepackage{graphicx}
\usepackage{amssymb}
\usepackage{bm}

\begin{document}

\title{A Specific Case of Generalized Einstein-aether Theories}

\author{Xinhe Meng $^{1, 3}$}\email{xhm@nankai.edu.cn}
\author{Xiaolong Du$^{1, 2}$}\email{duxiaolong@mail.nankai.edu.cn}

\affiliation{$^{1}$Department of Physics, Nankai University, Tianjin
300071, P.R.China} \affiliation{$^{2}$Department of Physics, Lanzhou
University, Lanzhou 730000, P.R.China} \affiliation{$^{3}$ Kavli
Institute of Theoretical Physics China,\\CAS, Beijing 100190,
China.\vspace{1cm}}
\date{\today}

\begin{abstract}
With the dark energy phenomena explored over a decade, in this
present work we discuss a specific case of the generalized
Einstein-aether theories, in which the modified Friedmann Equation
is similar to that in the Dvali-Gabadadze-Porrati (DGP) brane world
model. We compute the joint statistic constraints on model
parameters in this specific case by using the recent type Ia
supernovae (SNe Ia) data, the Cosmic Microwave Background (CMB)
shift parameter data, and the Baryonic Acoustic Oscillations (BAOs)
data traced by the Sloan Digital Sky Survey (SDSS). Furthermore, we
analyze other constrains from the observational Hubble parameter
data (OHD). The comparison with the standard cosmological model
(cosmological constant $\Lambda$ Cold Dark Matter ($\Lambda$CDM)
model) is clearly shown; also we comment on the interesting relation
between the coupling constant $M$ in this model and the special
accelerate scale in the MOdified Newtonian Dynamics (MOND) model
initially given by Milgrom with the hope for interpreting the galaxy
rotation curves without introducing mysterious dark matter.
\end{abstract}

\maketitle

\section{Introduction}\label{intro}
The independent discovery respectively in $1998$ and $1999$ that the
current universe's expansion is actually speeding up rather than
previously thought slowing-down due to the evolution dynamics
dominated by cosmic matter component (mainly the speculated
mysterious cold dark matter)~\cite{Riess:1998cb} has been an amazing
result. To account for that cosmic accelerating expansion, together
with other astrophysics observations, such as CMB, large scale
structure survey, like the SDSS, and the the universe age or Hubble
constant measurements, a so coined dark-energy component (even more
puzzling than the dark matter concept) with enough negative pressure
has been hypothesized. According to the Wilkinson Microwave
Anisotropy Probe (WMAP) 7-year data-set
analysis~\cite{Jarosik:2010iu} it makes up about $72.8\%$ of the
universe's contents. However, dark energy maybe the most mysterious
component of the universe hitherto as envisioned, we know little
about what it is and its nature. Over the past decade there have
been many theoretical models for mimicking the dark energy
behaviors, such as the simplest (just) cosmological constant and the
popular quintessence models~\cite{quintessence}. An alternatively
instructive idea is that the general theory of relativity may fail
to describe the universe's evolution on very large cosmic scales. Some
attempts have been made at modifying the standard general relativity
such as the $f(R)$ extended gravity models~\cite{fR1}, string theory
inspired cosmology models, brane cosmology and the holographic
principle applicable to the universe's evolution modelings. In this
present work we concentrate on the generalized Einstein-aether
theories as proposed by T. G. Zlosnik, P. G. Ferreira and G. D.
Starkman~\cite {Zlosnik:2006zu,Zlosnik:2007bu}, which is a
generalization of the Einstein-aether theory developed by Ted
Jacobson and David Mattingly~\cite {Jacobson:2000xp,Jacobson:2004ts}
with a free general function ${\cal F}({\cal K}) $ for model
buildings.

Arrangement for this paper is as follows. In the next section,
Section~\ref{aether}, we briefly review the framework of generalized
Einstein-aether theory by providing its basic equations for our later use. In
Section~\ref{FriedmannEqns}, we take a specific form of the ${\cal
F}({\cal K})$ function allowed and discuss the corresponding
modified Friedmann equations. In Section~\ref{data} followed, we
describe how to employ the observational data sets used for joint
statistics analysis in details with the hope that this clear
development can be useful to the related astrophysics and cosmology
community. In Section~\ref{result} by figures and tables, we show
our results compared with the currently standard cosmology
$\Lambda$CDM model. The last section, Section~\ref{conclusion}
contains our conclusions and discussions.

\section{Generalized Einstein-Aether Theories}\label{aether}
In the early history for  modern physics, the concept aether is
considered to be a physical medium homogenously occupying every
point in our universe. It determines a special rest reference frame,
in which everything has absolute relative velocity respect to it.
That suits for Newtonian dynamics very well. Later, this puzzling
concept is rejected by Einstein's relativity with mainly optics
experiments. By saying ``aether" framework, a term in this present
paper, we do not mean a mechanical medium naively, but rather a
locally preferred state resting for each point of the spacetime in
our physical evolutionary universe, determined by some hitherto
unknown physics or its physical state is to be specified by some
physical conditions with its environment. Some people even argue
that the smoothly distributed CMB everywhere may be regarded as a
modern version of aether. Einstein-aether theories were popularized
by Gasperini in a series of papers~\cite{Gasperini}. A vector-tensor
theory is suggested by Ted Jacobson and David Mattingly~\cite
{Jacobson:2000xp,Jacobson:2004ts}, where in addition to the metric
tensor field of general relativity this theory also contains a time-like
unit vector field which picks out a preferred frame at each point in the
spacetime. Then it is generalized by T. G. Zlosnik, P. G. Ferreira and
G. D. Starkman~\cite{Zlosnik:2006zu,Zlosnik:2007bu}.

The action of this theory with the normal Einstein-Hilbert part
action can be written in the form below
\begin{equation}
S=\int d^{4}x \sqrt{-g} \left[\frac{R}{16 \pi G}+{\cal L}_{A}+{\cal L}_{M}\right],
\label{S}
\end{equation}
where ${\cal L}_{A}$ is the vector field Lagrange density while
${\cal L}_{M}$ denotes the Lagrange density for all other matter
fields. The Lagrange density for the vector part consists of terms
quadratic in the field and its derivatives~\cite {Zlosnik:2006zu}:
\begin{eqnarray}
{\cal L}_{A}&=&\frac{M^{2}}{16 \pi G} {\cal F}({\cal K})+\frac{1}{16 \pi G} \lambda
(A^{\alpha}A_{\alpha}+1)\nonumber
\\
{\cal K}&=&M^{-2} {\cal K}^{\alpha \beta}_{\phantom{\alpha \beta} \gamma \sigma}
\nabla_{\alpha}A^{\gamma}\nabla_{\beta}A^{\sigma}\nonumber
\\
{\cal K}^{\alpha \beta}_{\phantom{\alpha \beta} \gamma \sigma}&=&c_{1}g^{\alpha
\beta}g_{\gamma \sigma}+c_{2}\delta^{\alpha}_{\gamma}\delta^{\beta}_{\sigma}+
c_{3}\delta^{\alpha}_{\sigma}\delta^{\beta}_{\gamma},
\label{L_A}
\end{eqnarray}
where $c_{i}$ are dimensionless constants and the coupling constant
$M$ has the dimension of mass. The ${\cal F}({\cal K})$ is a free
function that we do not know a priori, and the $\lambda$ is a
Lagrange multiplier that enforces the unit constraint for the
time-like vector field. In some papers an additional term
$c_{4}A^{\alpha}A^{\beta}g_{\gamma\sigma}$ is also included in the
expression for ${\cal K}^{\alpha\beta}_{\phantom{\alpha \beta}
\gamma \sigma}$~\cite{Jacobson:2004ts}.

We choose the inverse metric tensor $g^{\alpha\beta}$ and the
contravariant vector field $A^{\beta}$ to be our dynamic degrees of
freedom. Field equations from varying the action (\ref{S}) with
respect to $g^{\alpha\beta}$ and $A^{\beta}$ respectively are given
by
\begin{eqnarray}
G_{\alpha\beta}&=&\tilde {T}_{\alpha\beta}+8 \pi G T^{matter}_{\alpha\beta}
\label{eqn_g}
\\
\nabla_{\alpha}({\cal F}' J^{\alpha}_{\phantom{\alpha}\beta})&=& 2\lambda A_{\beta},
\label{eqn_A}
\end{eqnarray}
where $\tilde {T}_{\alpha\beta}$ is the stress-energy tensor for the vector field
and
\begin{equation}
{\cal F}'=\frac{d{\cal F}}{d{\cal K}}, \quad \quad
J^{\alpha}_{\phantom{\alpha}\sigma}=2 {\cal K}^{\alpha
\beta}_{\phantom{\alpha\beta} \sigma
\gamma}\nabla_{\beta}A^{\gamma}. \label{F'_J}
\end{equation}
For the choice (\ref{L_A}), $\tilde {T}_{\alpha\beta}$ is given by~\cite{Zlosnik:2006zu}
\begin{equation}
\tilde {T}_{\alpha\beta}=\frac{1}{2}\nabla_{\sigma} \left[{\cal F}'
(J_{(\alpha}^{\phantom{\alpha}\sigma}A_{\beta)}-J^{\sigma}_{\phantom
{\sigma}(\alpha}A_{\beta)}-J_{(\alpha\beta)}A^{\sigma})\right]-{\cal F}'
Y_{(\alpha\beta)}+\frac{1}{2}g_{\alpha\beta}M^2{\cal F}+\lambda A_{\alpha}
A_{\beta},
\label{Tab}
\end{equation}
where the sub $(ab)$ means symmetric with respect to the indices
involved and
\begin{equation}
Y_{\alpha\beta}=-c_{1}\left[(\nabla_{\nu}A_{\alpha})(\nabla^{\nu}A_{\beta})-
(\nabla_{\alpha}A_{\nu})(\nabla_{\beta}A^{\nu})\right].
\label{Yab}
\end{equation}
In addition, the constraint that {\bf A} is a time-like unit vector field gives
$A^{\alpha}A_{\alpha}=-1$.

\section{Modified Friedmann Equations}\label{FriedmannEqns}
Now we consider the case of a homogeneous and isotropic universe as
preferred by the WMAP observations, which can be described by the
Friedmann-Robertson-Walker metric
\begin{equation}
ds^2=-dt^2+a^2(t)\left(\frac{1}{1-kr^2}dr^2+r^2d\Omega^2\right),
\label{FRW}
\end{equation}
where $k$ is the curvature parameter. In such a case, the vector
must respect the spatial homogeneity and isotropy of the universe at
large scales. Thus the only component non-vanishing is the time-like
component. Using the constraint $A^{\alpha}A_{\alpha}=-1$, we can
get
\begin{equation}
A^{\alpha}=(1,0,0,0).
\label{A}
\end{equation}
We take the matter component as a perfect fluid, so its
energy-momentum tensor is of the form
\begin{equation}
T^{matter}_{\alpha\beta}=\rho U_{\alpha}U_{\beta}+p (U_{\alpha}U_{\beta}+g_{
\alpha\beta}),
\label{fluid}
\end{equation}
where $U_{\alpha}$ is the fluid four-velocity. By using (\ref{FRW})
and (\ref{A}), ${\cal K}$ can be simplified as
\begin{eqnarray}
{\cal K}&=&M^{-2}(c_{1}g^{\alpha\beta}g_{\gamma \sigma}+c_{2}\delta^{\alpha}_{\gamma}
\delta^{\beta}_{\sigma}+c_{3}\delta^{\alpha}_{\sigma}\delta^{\beta}_{\gamma})
\nonumber
\\
&=&3\alpha\frac{H^2}{M^2},
\label{Ksim}
\end{eqnarray}
where the coefficient $\alpha=c_{1}+3c_{2}+c_{3}$ and the Hubble
parameter $H\equiv\dot{a}/a$. It is easy to find by calculation that
the stress-energy tensor for the vector field also takes the form of
a perfect fluid, with an energy density given by (also see in \cite
{Zlosnik:2007bu})
\begin{equation}
\rho_{A}=3\alpha H^2({\cal F}'-\frac{\cal F}{2 \cal K})
\label{rho_A}
\end{equation}
and a pressure as
\begin{equation}
p_{A}=3\alpha H^2(-\frac{2}{3}{\cal F}'+\frac{\cal F}{2\cal K})-\alpha\dot{
\cal F}'H-\alpha{\cal F}'\frac{\ddot{a}}{a}.
\label{p_A}
\end{equation}
We can check that the vector field part's contributions obey the
cosmological energy conservation relation
$\dot{\rho}_A+3H(\rho_A+p_{A})=0$. A simple case has been discussed
by Sean M. Carroll and Eugene A. Lim in
reference~\cite{Carroll:2004ai}. Now we show that this conservation
relation is applicable to an arbitrary form of ${\cal F}({\cal K})$.

By taking eqs.(\ref{FRW})-(\ref{Ksim}) into field equations
(\ref{eqn_g}) and (\ref{eqn_A}), the modified Friedmann equations
can be derived (see also in \cite{Zlosnik:2007bu}):
\begin{eqnarray}
(1-\alpha{\cal F}'+\frac{1}{2}\frac{\alpha{\cal F}}{\cal K})H^2+\frac{k}{a^2}&=&\frac{
8\pi G}{3}\rho
\label{Friedmann1}
\\
\frac{d}{dt}(-2H+\alpha{\cal F}'H)+\frac{2k}{a^2}&=&8\pi G(\rho+p).
\label{Friedmann2}
\end{eqnarray}
In order to see what has been modified, we list the standard
Friedmann equations in the $\Lambda$CDM model below for comparison:
\begin{eqnarray}
H^2+\frac{k}{a^2}-\frac{\Lambda}{3}&=&\frac{8\pi G}{3}\rho
\label{Friedmann0_1}
\\
-2\frac{dH}{dt}+\frac{2k}{a^2}&=&8\pi G(\rho+p).
\label{Friedmann0_2}
\end{eqnarray}
We can see that a few terms involving ${\cal F}({\cal K})$ and its
first order derivative are present, which may imply that an
effective term involving cosmological ``constant" or an effective
cosmological ``constant" (the effective vacuum energy for the
universe) can be from the vector field's contributions, the mysterious
``aether".

In their interesting works~\cite{Zlosnik:2007bu}
and~\cite{Zuntz:2010jp}, a class of theories with ${\cal F}({\cal
K})=\gamma(-{\cal K})^n$ have been discussed. Noting that
$\alpha\leqslant0$~\cite{Lim:2004js}, and ${\cal K}$ is negative
there. It has been shown that the scale $M\thicksim H_{0}$ and for
appropriate parameters the generalized Einstein-aether theories can
lead to a late-time acceleration of the universe's
expansion (for example, the $n=0$ case is just corresponding to the
$\Lambda$CDM model). For that form of ${\cal F}({\cal K})$, the first
modified Friedmann equation (\ref{Friedmann1}) becomes~\cite{Zlosnik:2007bu}
\begin{equation}
\left[1+\epsilon \left(\frac{H}{M}\right)^{2(n-1)}\right]H^2+\frac{k}{a^2}=\frac{8
\pi G}{3}\rho,
\label{Friedmann1_2}
\end{equation}
where $\epsilon=-(2n-1)\gamma(-2\alpha)^n/6$. We can see if
$n=\frac{1}{2}$, $\epsilon=0$ and the modified terms disappear.
Thus, there will not be a modified term proportional to $H$.
However, what about other forms of ${\cal F}({\cal K})$?

In the following part of this paper we consider a specific case, in which we take
\begin{equation}
{\cal F}({\cal K})=\beta\sqrt{-{\cal K}}+\sqrt{\frac{2{\cal K}}{\alpha}}\ln(-{\cal K}),
\label{F(K)}
\end{equation}
where $\beta$ is a constant. Taking equation (\ref{F(K)}) into (\ref{Friedmann1}),
Equation (\ref{Friedmann1}) then becomes
\begin{equation}
H^2-MH+\frac{k}{a^2}=\frac{8\pi G}{3}\rho.
\label{Friedmann1_1}
\end{equation}
where we have used ${\cal K}=3\alpha H^2/M^2$ as in given equation
(\ref{Ksim}). For $k=0$, equation (\ref{Friedmann1_1}) is almost the
same as that in DGP brane world model~\cite
{Dvali:2000hr,Dvali:2000xg,Deffayet:2000uy}. For $k\neq0$, it will
be a little different. It is still unknown whether there is any
relation between these two theories.

It is easy to figure out that when $\rho\rightarrow0$,
$H\rightarrow M$ for the evolutionary universe whose geometry is almost
flat at the late stage as indicated from the WMAP observations now. That
is to say at late period of the universe's evolution when
$\rho\varpropto a^{-3} \rightarrow 0$, the universe will keep its
accelerating expansion due to the existence of the aether field and
an interesting scale appears $M\rightarrow H$ then.

Furthermore, we can calculate the effective state parameter for the
vector field part and the deceleration parameter for our choice of
the function ${\cal F(\cal K)}$:
\begin{eqnarray}
w_{A}&\equiv&\frac{p_{A}}{\rho_{A}}=-\frac{\dot{H}}{3H^2}-1
\label{w_A}
\\
q&\equiv& -\frac{a\ddot{a}}{\dot{a}^2}=-\frac{\dot{H}}{H^2}-1=3(w_A+1)-1.
\label{q}
\end{eqnarray}
From the (\ref{q}) we know directly that to explain the speeding up
of the universe's expansion as implied by lots of astrophysics
observations, the effective state parameter for the vector field
part's contributions today must be smaller than $-\frac{2}{3}$
(instead of the $-\frac{1}{3}$ as given directly from the standard Friedmann
equations for the $\Lambda$CDM model).

\section{Current Observational Data}\label{data}

\subsection{Type Ia Supernovae}\label{supernovae}
The observations of Type Ia supernovae (SNe Ia) provide an excellent tool for probing the
expansion history of the universe. Because all type Ia supernovae explode at about the
same mass, their absolute magnitudes are considered to be all the same ($M\thickapprox
-19.3\pm0.3$). This makes them very useful as standard candles. The observation of
supernovae measure essentially the apparent magnitude $m$. The theoretical distance modulus
is defined as
\begin{equation}
\mu_{th}=m_{th}-M=5\log_{10}D_{L}(z)+\mu_{0},
\label{m}
\end{equation}
where $D_{L}(z)\equiv H_{0}d_{L}(z)$ is the dimensionless luminosity and $\mu_{0}=42.38-5
\log_{10}h$. Here $h$ is the dimensionless Hubble parameter toady. $D_{L}(z)$ is given by
\begin{equation}
D_{L}(z)=\frac{1+z}{\sqrt{\Omega_{k}}}\sinh \left[H_{0}\sqrt{\Omega_{k}}\int^{z}
_{0}\frac{dz'}{H(z)}\right],
\label{D_L}
\end{equation}
where $\Omega_{k}$ is the fractional curvature density today. In this paper we use the
Union2 data set consisting of 557 supernovae. The corresponding $\chi^2_{SN}$ function
to be minimized is
\begin{equation}
\chi^2_{SN}=\sum_{i=1}^{557}\left[\frac{\mu_{obs}(z_{i})-\mu_{th}(z_{i};{\bf\theta})}
{\sigma_{i}}\right]^2,
\label{chi_SN}
\end{equation}
where ${\bf\theta}$ denotes the model parameters. The minimization with respect to
$\mu_0$ can be made trivially by expanding $\chi^2_{SN}$ with respect to $\mu_0$ as
\cite{Nesseris:2005ur}
\begin{equation}
\chi^2_{SN}({\bf\theta})=A-2\mu_{0}B+\mu_{0}^2C,
\label{chi_ex}
\end{equation}
where
\begin{eqnarray}
A({\bf \theta})&=&\sum_{i=1}^{557}\frac{[\mu_{obs}(z_{i})-\mu_{th}(z_{i};
\mu_0=0,{\bf\theta})]^2}{\sigma_{i}^2}
\label{A(theta)}
\\
B({\bf \theta})&=&\sum_{i=1}^{557}\frac{\mu_{obs}(z_{i})-\mu_{th}(z_{i};
\mu_0=0,{\bf\theta})}{\sigma_{i}^2}
\label{B(theta)}
\\
C&=&\sum_{i=1}^{557}\frac{1}{\sigma_{i}^2}.
\label{C}
\end{eqnarray}
Equation (\ref{chi_ex}) has a minimum for $\mu_{0}=B/C$ at
\begin{equation}
\tilde{\chi}^2_{SN}({\bf \theta})=A({\bf \theta})-\frac{B^2({\bf \theta})}
{C}.
\label{chi_SN_1}
\end{equation}
Thus instead of minimizing $\chi^2_{SN}$ we can minimize $\tilde{\chi}^2_{SN}$ which
is independent of $\mu_{0}$.

\subsection{Cosmic Microwave Background and Baryonic Acoustic Oscillations}
\label{CMB_BAOs}
In addition to the type Ia supernovae data, we use the Cosmic Microwave Background (CMB)
shift parameter and the Baryonic Acoustic Oscillations to compute the joint constraints.
The shift parameter ${\cal R}$ is defined in~\cite{Bond:1997wr} as
\begin{equation}
{\cal R}\equiv\sqrt{\Omega_{m}H^2_{0}}(1+z_{*})D_{A}(z_{*}),
\label{R}
\end{equation}
where $z_{*}$ is the redshift of recombination and $D_{A}(z)$ is the proper angular
diameter distance:
\begin{equation}
D_{A}(z)=\frac{1}{H_{0}(1+z)\sqrt{\Omega_{k}}}\sinh \left[H_{0}
\sqrt{\Omega_{k}}\int^{z}_{0}\frac{dz'}{H(z)}\right].
\label{D_A}
\end{equation}
The seven-year WMAP results~\cite{Komatsu:2010fb} have updated the redshift of recombination
$z_{*}=1091.3$ and the shift parameter ${\cal R}=1.725\pm0.018$. The $\chi^2$ for CMB shift is
\begin{equation}
\chi^2_{CMB}({\bf \theta})=\frac{[{\cal R}({\bf \theta})-1.725]^2}{0.018^2}
\label{chi_CMB}
\end{equation}

Another constraint is from the Baryonic Acoustic Oscillations (BAOs) traced by the Sloan
Digital Sky Survey (SDSS). In this paper we use only one node at $z=0.35$. The distance
parameter ${\cal A}$ is defined as\cite{Eisenstein:2005su}
\begin{equation}
{\cal A}\equiv D_{V}(0.35)\frac{\sqrt{\Omega_{m}H_{0}^2}}{0.35},
\label{AA}
\end{equation}
where $D_{V}$ is the effective distance
\begin{equation}
D_{V}(z)=\left[(1+z)^2 D_{A}^2(z)\frac{z}{H(z)}\right]^{\frac{1}{3}}.
\label{D_V}
\end{equation}
The value of ${\cal A}$ is given in~\cite{Eisenstein:2005su}: ${\cal A}=0.469\pm0.017$.
Thus the $\chi^2$ for BAO is
\begin{equation}
\chi^2_{BAO}({\bf \theta})=\frac{[{\cal A}({\bf \theta})-0.469]^2}{0.017^2}.
\label{chi_BAO}
\end{equation}
To compute the joint constraints, we add these $\chi^2$ functions together:
\begin{equation}
\chi^2=\tilde{\chi}_{SN}^2+\chi_{CMB}^2+\chi_{BAO}^2.
\label{chi_t}
\end{equation}

\subsection{Observational Hubble Parameter Data}\label{OHD}
There are two major methods of independent observational $H(z)$
measurement, which are called the ``differential age method" and the
``radial BAO size method". The details can be found
in~\cite{Zhang:2010ic}. In that paper, Tong-Jie Zhang and Cong Ma
summarize the up-to-date observational Hubble parameter data (OHD).
See below in Table~\ref {OHD_data}. The data points at
$z=0.24$ and $z=0.43$ are derived from the ``radial BAO size
method", while the others are derived from the ``differential age
method" as named.

The $\chi^2$ for OHD is
\begin{equation}
\chi_{OHD}^2=\sum_{i=0}^{13}\frac{[H_{0}E(z_{i})-H_{obs}(z_{i})]^2}{\sigma_{i}^2},
\label{chi_OHD}
\end{equation}
where $E(z)\equiv H(z)/H_{0}$ is independent of $H_{0}$. Using the same trick mentioned
before, the minimization with respect to $H_0$ can be made trivially by expanding $\chi^2
_{OHD}$ with respect to $H_0$ as
\begin{equation}
\chi^2_{OHD}({\bf\theta})=H_{0}^2 A-2H_{0}B+C,
\label{chi_OHD_ex}
\end{equation}
where
\begin{eqnarray}
A&=&\sum_{i=1}^{13}\frac{E^2(z_i)}{\sigma_i^2}
\label{A_OHD}
\\
B&=&\sum_{i=1}^{13}\frac{E(z_i)H_{obs}(z_i)}{\sigma_i^2}
\label{B_OHD}
\\
C&=&\sum_{i=1}^{13}\frac{H_{obs}^2(z_i)}{\sigma_i^2}.
\label{C_OHD}
\end{eqnarray}
Equation (\ref{chi_OHD_ex}) has a minimum for $H_{0}=B/A$ at
\begin{equation}
\tilde{\chi}^2_{OHD}=-\frac{B^2}{A}+C.
\label{chi_OHD_1}
\end{equation}
Thus, instead of minimizing $\chi^2_{OHD}$ we can minimize
$\tilde{\chi}^2_{OHD}$ which is independent of $H_{0}$. From
Table~\ref{OHD_data}, we can see the errors of OHD (data sets)
listed are relatively larger. So we do analysis only with the OHD
(data sets) separately with the hope that we can obtain more
accuracy OHD in the near future.

\begin{table}[htbp]
\caption{The set of available observational $H(z)$ data}
\label{OHD_data}
\begin{tabular}{ccc}
\hline\hline
        $~~~~~~z~~~~~~$  & $~~~~~~H(z)\pm 1\sigma~~~~~~$ & ~~~~~~References~~~~~~\\
                         & km s$^{-1}$Mpc$^{-1}$         &      \\
\hline
        0.09 & $69\pm12$  & \cite{Jimenez:2003iv, Stern:2009ep} \\
        0.17 & $83\pm8$   & \cite{Stern:2009ep}  \\
        0.24 & $79.69\pm2.65$& \cite{Gaztanaga:2008xz}  \\
        0.27 & $77\pm14$  & \cite{Stern:2009ep}    \\
        0.4  & $95\pm17$  & \cite{Stern:2009ep}    \\
        0.43 & $86.45\pm3.68$& \cite{Gaztanaga:2008xz}  \\
        0.48 & $97\pm62$  & \cite{Stern:2009ep}    \\
        0.88 & $90\pm40$  & \cite{Stern:2009ep}    \\
        0.9  & $117\pm23$ & \cite{Stern:2009ep}    \\
        1.3  & $168\pm17$ & \cite{Stern:2009ep}    \\
        1.43 & $177\pm18$ & \cite{Stern:2009ep}    \\
        1.53 & $140\pm14$ & \cite{Stern:2009ep}    \\
        1.75 & $202\pm40$ & \cite{Stern:2009ep}    \\
\hline\hline
\end{tabular}
\end{table}

\section{Results}\label{result}
For the $\Lambda$CDM model, the general expression for the expansion
relation can be directly written out as
\begin{equation}
H(z)=H_{0}\sqrt{\Omega_{\Lambda}+\Omega_{k}(1+z)^2+\Omega_{m}(1+z)^3+\Omega_{R}(1+z)^4},
\label{H_LCDM}
\end{equation}
where $\Omega_{\Lambda}$, $\Omega_{k}$, $\Omega_{m}$, $\Omega_{R}$ are the fractional
density of vacuum, curvature, matter and radiation today, respectively.
\begin{equation}
\Omega_{\Lambda}\equiv\frac{8\pi \Lambda}{3H_{0}^2},\quad\quad
\Omega_{k}\equiv\frac{-k}{a_{0}^2H_{0}^2},\quad\quad
\Omega_{m}\equiv\frac{8\pi \rho_{m}}{3H_{0}^2},\quad\quad
\Omega_{R}\equiv\frac{8\pi \rho_{R}}{3H_{0}^2}\quad\quad.
\end{equation}
In this paper, we fix $\Omega_{R}=\Omega_{\gamma}(1+0.2271N_{eff})$ and take the present
photon density parameter $\Omega_{\gamma}=2.469\times10^{-5}h^{-2}$ (for $T_{cmb}=2.725K$)
and the effective number of neutrino species at its standard value 3.04~\cite{Komatsu:2010fb}.
We also use the prior $h=74.2\pm3.6$ given in~\cite{Riess:2009pu}. So there are only two
independent parameters $\Omega_{\Lambda}$ and $\Omega_{m}$. The parameter $\Omega_{k}$ can
be expressed by the others:
\begin{equation}
\Omega_{k}=1-\Omega_{\Lambda}-\Omega_{m}-\Omega_{R}.
\label{Omega_k}
\end{equation}

For the Einstein-aether theory with our choice of the free function
${\cal F}({\cal K})$, we can solve $H(z)$ from (\ref{Friedmann1_1}):
\begin{equation}
H(z)=H_{0}\left[\frac{\Omega_{A}}{2}+\sqrt{\frac{\Omega_{A}^2}{4}+\Omega_{k}(1+z)^2+
\Omega_{m}(1+z)^3+\Omega_{R}(1+z)^4}\right],
\label{H_Aether}
\end{equation}
where we define
\begin{equation}
\Omega_{A}\equiv\frac{M}{H_{0}}.
\end{equation}
There are also only two parameters $\Omega_{A}$ and $\Omega_{m}$. Similarly, $\Omega_{k}$
can be expressed as (\ref{Omega_k}), in which $\Omega_{\Lambda}$ needs to be replaced by
$\Omega_{A}$.

\begin{figure}[htbp]
\begin{center}
\includegraphics[width=3in]{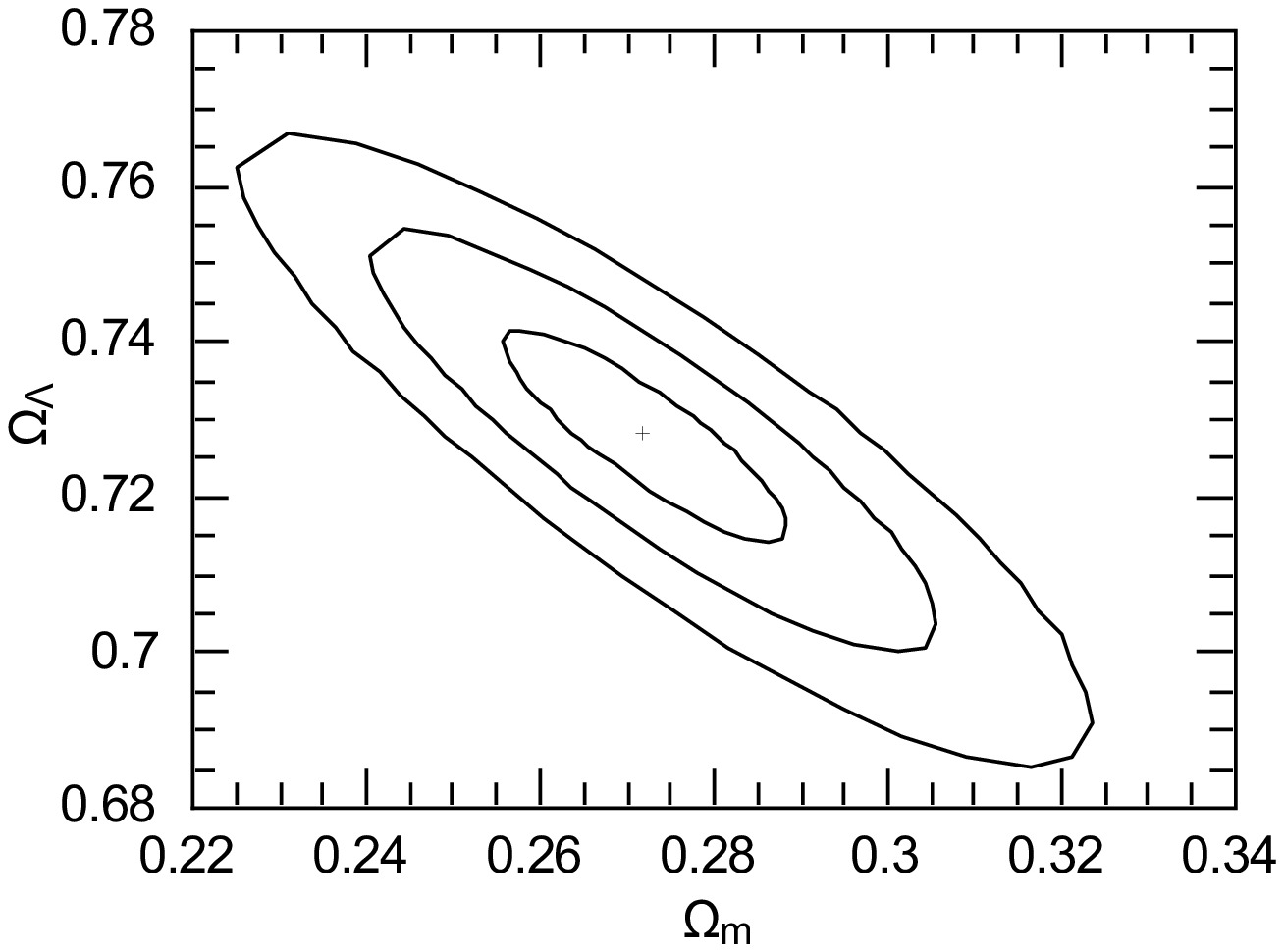}
\includegraphics[width=3in]{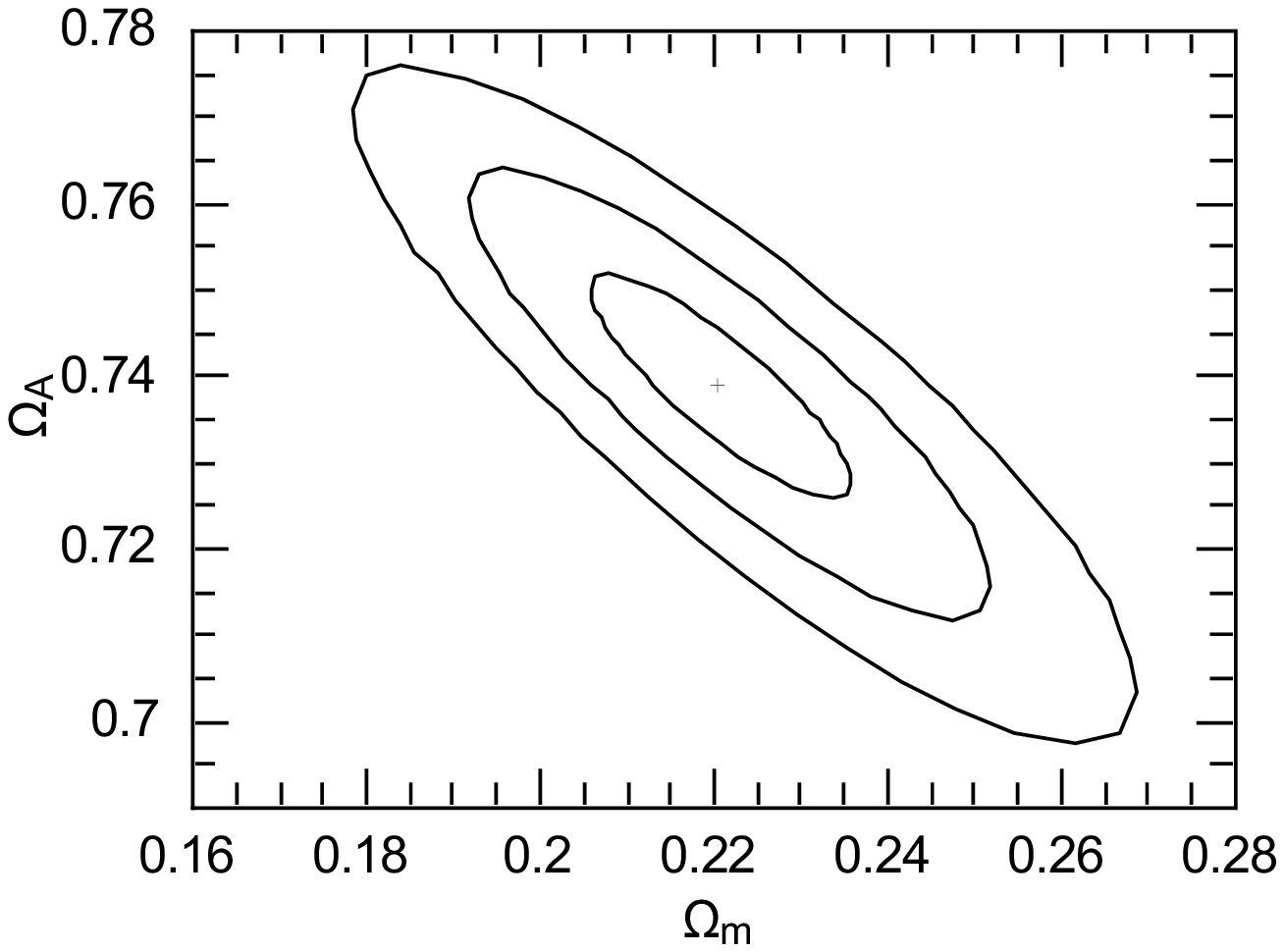}
\end{center}
\caption{Joint constraints from SNe Ia, CMB shift and BAO. The
$1\sigma$, $2\sigma$, $3\sigma$ confidence interval contours of
$\Omega_{m}$ and $\Omega_{\Lambda}$ (or $\Omega_{A}$ for the
Einstein-aether theory case) in $\Lambda$CDM model (left) and the
Einstein-aether theory case(right).} \label{pic1}
\end{figure}

Firstly, we compute the combined constraints from SNe Ia, CMB shift
and BAO data sets.  The results are shown in Table~\ref{best_fit}
and Figure~\ref{pic1}. The best-fit values for parameters of the
$\Lambda$CDM model are consistent with the results given by WMAP
7~\cite{Jarosik:2010iu}. It gives a nearly flat universe geometry
with a tiny $\Omega_{k}=4\times10^{-5}$. For the Einstein-aether
theory case as we choose, the best-fit $\Omega_{m}$ is a little bit
smaller than that in $\Lambda$CDM model, but it gives a larger
$\Omega_{k}=0.04$. The $\Lambda$CDM model fits better to the data
sets as its $\chi_{min}^2$ is $26.182$ smaller than that of
Einstein-aether theory case as shown.

\begin{figure}[htbp]
\begin{center}
\includegraphics[width=3in]{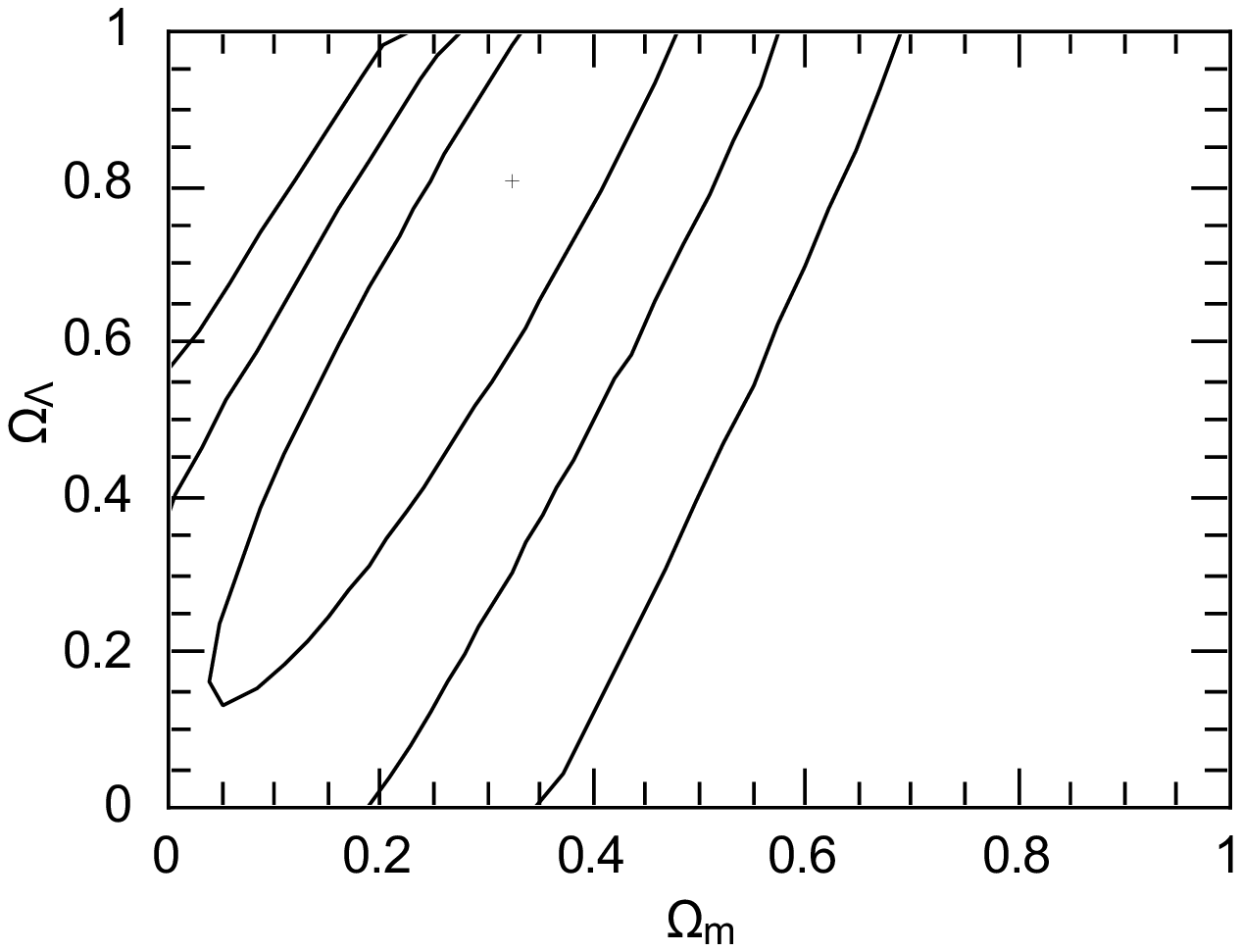}
\includegraphics[width=3in]{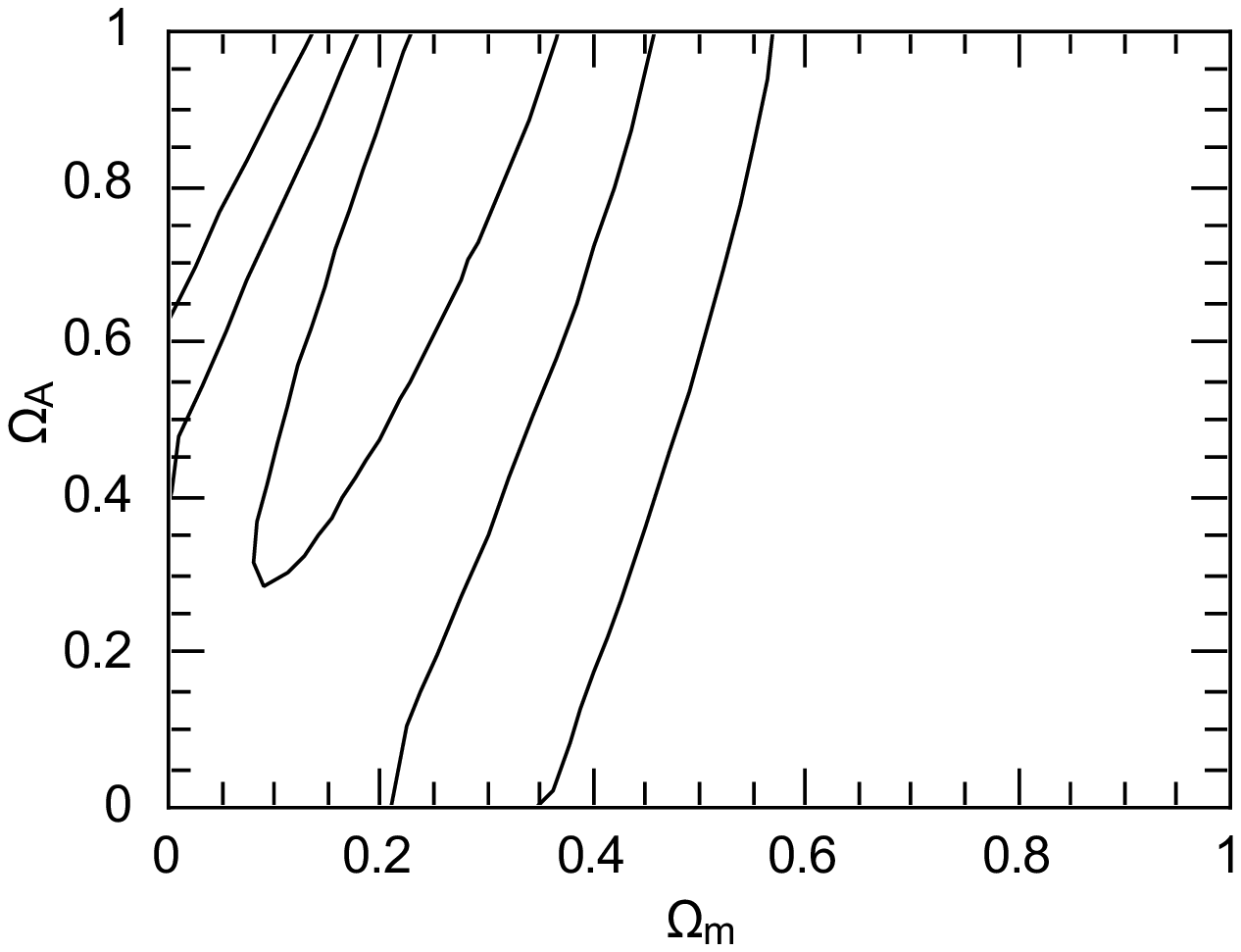}
\end{center}
\caption{Constraints from OHD. The $1\sigma$, $2\sigma$, $3\sigma$ confidence
interval contours of $\Omega_{m}$ and $\Omega_{\Lambda}$ (or $\Omega_{A}$ for
Einstein-aether theory case) in $\Lambda$CDM model (left) and Einstein-aether
theory case (right).}
\label{pic2}
\end{figure}

Then, we do similar analysis with the OHD (data sets). The results
are shown in Table~\ref{best_fit} and Figure~\ref{pic2} as well.
This time, we can see that the Einstein-aether theory case fits a
little better to the data sets, and the difference of the
$\chi_{min}^2$ is very small now.

Noting the similarity of our modified Friedmann equation to that in
DGP brane world model and data analysis being done to DGP model~\cite
{Li:2009jx,Wan:2007pm}, these result are rather natural, because
best-fit $\Omega_{k}$ is very small.

\begin{table}[htbp]
\caption{Best-fit parameters for $\Lambda$CDM model and Einstein-aether theory}
\label{best_fit}
\begin{tabular}{cccc}
\hline\hline
                                &~~~~~~Model~~~~~~&~~~~~~Best-fit parameters~~~~~~&~~~~~~$\chi^2_{min}$~~~~~~\\
\hline
        SN Ia-CMB shift-BAO    &  $\Lambda$CDM            &  $\Omega_{m}=0.272$       & 542.693 \\
                                &                          &  $\Omega_{\Lambda}=0.728$ &         \\
                                &  Einstein-aether theory  &  $\Omega_{m}=0.220$       & 568.875 \\
                                &                          &  $\Omega_{A}=0.739$       &         \\
\hline
        OHD                     &  $\Lambda$CDM            &  $\Omega_{m}=0.322$       & 8.046   \\
                                &                          &  $\Omega_{\Lambda}=0.807$ &         \\
                                &  Einstein-aether theory  &  $\Omega_{m}=0.301$       & 7.989   \\
                                &                          &  $\Omega_{A}=1.040$       &         \\
\hline\hline
\end{tabular}
\end{table}

\begin{figure}[htbp]
\begin{center}
\includegraphics[width=3in]{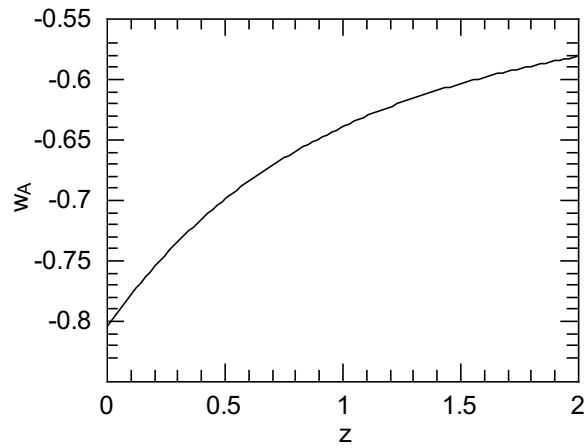}
\end{center}
\caption{The effective state parameter of the vector
field part's contribution.}
\label{pic3}
\end{figure}

\begin{figure}[htbp]
\begin{center}
\includegraphics[width=3in]{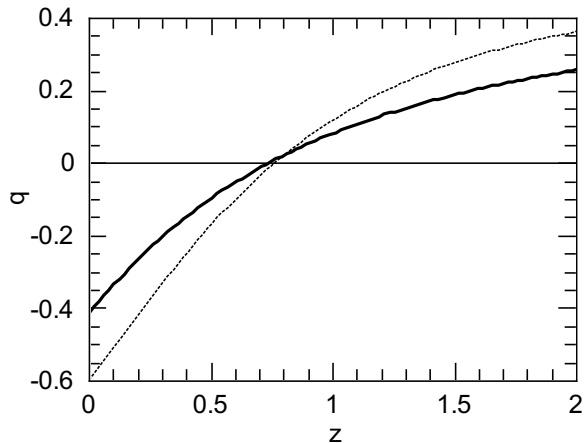}
\end{center}
\caption{The deceleration parameter. The thick solid line is the result
of the Einstein-aether theory, while the thin dashed line represents
$\Lambda$CDM model.}
\label{pic4}
\end{figure}

Using the results of combined analysis we also plot the effective
state parameter of the vector field part's contribution $w_{A}(z)$
(Figure~\ref{pic3}) and the corresponding deceleration parameter
$q(z)$ (Figure~\ref{pic4}), for demonstration.

\section{Conclusions and Discussions}\label{conclusion}
In this paper we only consider in details a specific case of
generalized Einstein-aether theories and compute the joint constraints from
observations such as SNe Ia, CMB shift, BAO data sets, and OHD
respectively. Even though we only investigate a specific case of generalized
Einstein-aether model, we already see that it has shown lots of
interesting features, by well fitting to the combined data sets of
the SNe Ia, CMB shift and BAO, as well as OHD  respectively, and
comparing with the $\Lambda$CDM model.

The observational Hubble parameter data we have possessed now are
relatively few and not so accurate. However, with the improving
quality of observational $H(z)$ data and more data points being
measured (more sample compiled hopefully), it will be certainly a
directly useful tool to test dark energy models and modified
theories of general relativity, as well as corresponding cosmology
models.

For the case we analyze in this paper, the modified Friedmann
equation is similar to that in DGP brane world model. It may be
caused by the special ${\cal F} ({\cal K})$ we choose. It looks
first rather strange if there is any possible relation between these
two theories, because the Einstein-aether theory is initially
proposed for possible Lorentz violation and preferred frame effects,
while the DGP brane world model considers a 3-brane embedded in a 5D
bulk space-time. However, it is not completely impossible now.

Moreover, it is clearly shown in this special model that
$M=\Omega_{A}H_{0}\thicksim H_{0}$, which is consistent with the
requirements of MOND limit~\cite
{Zlosnik:2006zu,Zlosnik:2007bu,Zuntz:2010jp}. However, further work
needs to be elaborated on the stability analysis of this specific
case we have considered.

For the last point of this present work we would like to make (but
not the least importance), we should emphasize especially that
theoretically we can not give for granted that the phenomenological
MOND theory can reproduce all the systematics of Rotational Curves
(RCs) observations, although the MOND model fits BETTER than the
$\Lambda$CDM based mass models. It is still far away to reproduce
the wide and far telling systematics of the spiral galaxies'  RCs
(and of the mass distribution in the corresponding galaxies)
\cite{sal}, so there will be lots of detail work to be done with any
modified gravity proposals at least on the galaxy scale.

\acknowledgements

During various stages for this present work, we would like to thank
professors Miao Li and Peng-Jie Zhang, other members of our group
Zhiyuan Ma, Yingbin Wang and Han Dong as well for beneficial
discussions. Thanks also to Jie Ren and Zhiyuan Ma for programming
helps. For this piece of research, we really appreciate professors
Glenn Starkman for useful suggestions by reading our manuscript, and
Paolo Salucci for helpful comments to the MOND phenomenology. This
work is partly supported by Natural Science Foundation of China
under Grant Nos.11075078 and 10675062, and by the project of
knowledge Innovation Program (PKIP) of Chinese Academy of Sciences
(CAS) under the grant No. KJCX2.YW.W10 through the KITPC
astrophysics and cosmology programmes where we have initiated this
present work.

\end{document}